# New intercalation superconductor Li$_x$(C$_6$H$_{16}$N$_2$)$_y$Fe$_{2-z}$Se$_2$ with a very long interlayer-spacing and $T_c$ = 38 K

Shohei Hosono, Takashi Noji, Takehiro Hatakeda, Takayuki Kawamata, Masatsune Kato, and Yoji Koike

*Department of Applied Physics, Tohoku University, 6-6-05 Aoba, Aramaki, Aoba-ku, Sendai 980-8579, Japan*

A new iron-based superconductor Li$_x$(C$_6$H$_{16}$N$_2$)$_y$Fe$_{2-z}$Se$_2$ with $T_c$ = 38 K has successfully been synthesized via intercalation of lithium and hexamethylenediamine into FeSe. The superconducting transition has been confirmed not only by the magnetic susceptibility measurements but also by the electrical resistivity ones. The interlayer spacing, namely, the distance between the neighboring Fe layers, $d$, is 16.225(5) Å and is the largest among those of the FeSe-based intercalation compounds. It has been found that the dependence of $T_c$ on $d$ in the FeSe-based intercalation superconductors appears domic.

The iron-based chalcogenide superconductor FeSe with the superconducting transition temperature, $T_c$, = 8 K[1] has a crystal structure composed of a simple stack of edge-sharing FeSe$_4$-tetrahedra layers which are analogous to FeAs$_4$-tetrahedra layers in the iron-based pnictide superconductors. Therefore, it is possible to intercalate atoms and molecules between the FeSe$_4$-tetrahedra layers. In fact, it has been revealed that potassium is intercalated into FeSe, so that $T_c$ of K$_x$Fe$_{2-y}$Se$_2$ is markedly enhanced to be ~ 31 K.[3] Moreover, it has been found that ammonia, NH$_3$, is intercalated into FeSe together with alkali or alkali-earth metals, so that the $c$-axis length is much enlarged and that $T_c$'s of $M_x$(NH$_3$)$_y$Fe$_{2-z}$Se$_2$ ($M$: alkali or alkali-earth metals) are further enhanced to be 31 - 46 K.[4-8] What is remarkable is that $T_c$ of K$_{0.28}$(NH$_3$)$_{0.47}$Fe$_{1.97}$Se$_2$ with the $c$-axis length of 15.56 Å is 44 K and is much higher than $T_c$ = 30 K of K$_{0.57}$(NH$_3$)$_{0.34}$Fe$_{2.02}$Se$_2$ with the $c$-axis length of 14.84 Å.[7] That is, it has turned out that $T_c$'s of various intercalation superconductors of $M_x$(NH$_3$)$_y$Fe$_{2-z}$Se$_2$ tend to increase with increasing $c$-axis length.[8] These results suggest that the enhancement of the two-dimensionality due to the increase of the $c$-axis length leads to the increase in $T_c$,[9] while the content of $M$ does not affect $T_c$ so much. The latter may be reasonable, taking into account the theoretical result that the density of states at the Fermi level is independent of the electron density in the two-dimensional free-electron model.

Following ammonia, pyridine, $C_5H_5N$, has also been reported to be intercalated into FeSe together with lithium.[10] It has been found that $T_c$ of $Li_x(C_5H_5N)_yFe_{2-z}Se_2$ with the $c$-axis length of 16.0549 Å is 45 K and that post-annealing of the intercalated sample drastically expands the $c$-axis length to 23.09648 Å and increases the superconducting shielding volume fraction, though $T_c$ does not change through the post-annealing.

Very recently, ethylenediamine (EDA), $C_2H_8N_2$, has also been reported to be intercalated into FeSe together with lithium or sodium.[11,12] It has been found that $T_c$ is 45 K both for $Li_x(C_2H_8N_2)_yFe_{2-z}Se_2$ with the $c$-axis length of 20.74(7) Å and for $Na_x(C_2H_8N_2)_yFe_{2-z}Se_2$ with the $c$-axis length of 21.9(1) Å. Accordingly, it seems that $T_c$ increases with increasing $c$-axis length and is saturated at ~ 45 K. It has attracted great interest whether $T_c$ increases or decreases with further increasing $c$-axis length.

Here, we report on the successful synthesis of a new superconductor $Li_x(C_6H_{16}N_2)_yFe_{2-z}Se_2$ with $T_c$ = 38 K and a very long $c$-axis length of 32.450(9) Å via intercalation of lithium and hexamethylenediamine (HMDA), $C_6H_{16}N_2$, into FeSe. The superconducting transition is observed not only in the magnetic susceptibility, $\chi$, measurements but also in the electrical resistivity, $\rho$, measurements. The relation between $T_c$ and the interlayer spacing, namely, the distance between the neighboring Fe layers, $d$, in the FeSe-based intercalation superconductors is discussed.

Polycrystalline samples of FeSe were prepared by the solid-state reaction method. Fe and Se powders were mixed in the molar ratio of Fe : Se = 1.02 : 1 and pressed into pellets. The pellets were sealed in an evacuated quartz tube and heated at 800°C for 40 h. The obtained pellets of FeSe were pulverized into powder to be used for the intercalation. Dissolved lithium metal in HMDA was intercalated into the powdery FeSe as follows. An appropriate amount of the powdery FeSe was placed in a beaker filled with 0.2 - 0.8 M solution of pure lithium metal dissolved in HMDA. The amount of FeSe was calculated in the molar ratio of Li : FeSe = 1 : 1 or 1 : 2. The reaction was carried out at 45°C for 5 days. The separation of the product from HMDA was easily made by the solidification of residual HMDA at the top cap of the beaker, keeping the temperature of the top cap of the beaker below the melting point of HMDA (42°C). All the process was performed in an argon-filled glove box. Both FeSe and the intercalated sample were characterized by powder x-ray diffraction using $CuK_\alpha$ radiation. For the intercalated sample, an airtight sample-holder was used. The diffraction patterns were analyzed using RIETAN-FP.[13] The chemical composition was determined by inductively coupled plasma optical emission spectrometry (ICP-OES).

Thermogravimetric (TG) measurements were performed in flowing gas of argon. In order to detect the superconducting transition, $\chi$ was measured using a superconducting quantum interference device (SQUID) magnetometer (Quantum Design, Model MPMS). Measurements of $\rho$ were also carried out by the standard dc four-probe method. For the $\rho$ measurements, the as-intercalated powdery sample was pressed into a pellet and sintered at 190°C for 15 h in an evacuated glass tube.

Figure 1 shows the powder x-ray diffraction pattern of the as-intercalated sample. The broad peak around $2\theta = 20°$ is due to the airtight sample-holder. Most of sharp Bragg peaks are due to the intercalation compound of $Li_x(C_6H_{16}N_2)_yFe_{2-z}Se_2$ and the host compound of FeSe, so that they are able to be indexed based on the $ThCr_2Si_2$-type and PbO-type structures, respectively. Therefore, it is found that lithium and HMDA are partially intercalated into FeSe, while there remains a non-intercalated region of FeSe in the as-intercalated sample. The lattice constants of Li$_x$(C$_6$H$_{16}$N$_2$)$_y$Fe$_{2-z}$Se$_2$ are calculated to be $a$ = 3.453(2) Å and $c$ = 32.450(9) Å in the as-intercalated sample. As shown in Fig. 1, the simulated x-ray diffraction pattern of a hypothetical compound of TmFe$_2$Se$_2$ with the same lattice constants, where the number of electrons of Tm in the unit cell is the same as that of Li(C$_6$H$_{16}$N$_2$), is in good agreement with the diffraction pattern of Li$_x$(C$_6$H$_{16}$N$_2$)$_y$Fe$_{2-z}$Se$_2$, suggesting that both lithium and HMDA are located in the crystal site similar to that of Tm in TmFe$_2$Se$_2$. Moreover, our previous results have revealed that the intercalation of only lithium into Fe(Se,Te) has neither effect on the superconductivity nor crystal structure.[14] Accordingly, it is concluded that not only lithium but also HMDA has been intercalated between the Se-Se layers of FeSe. As listed in Table I, the $a$-axis length is a little smaller than that of FeSe as in the case of Li$_x$(C$_2$H$_8$N$_2$)$_y$Fe$_{2-z}$Se$_2$, while the $c$-axis length is much larger than 20.74 Å of Li$_x$(C$_2$H$_8$N$_2$)$_y$Fe$_{2-z}$Se$_2$.[11] The latter is reasonable, because HMDA is a linear molecule longer than EDA. The $c$-axis length of Li$_x$(C$_6$H$_{16}$N$_2$)$_y$Fe$_{2-z}$Se$_2$ is the largest among those of the FeSe-based intercalation compounds.[12] Since the unit cell of Li$_x$(C$_6$H$_{16}$N$_2$)$_y$Fe$_{2-z}$Se$_2$ includes two Fe layers, $d$ is as large as 16.225(5) Å and is, of course, the largest among those of the FeSe-based intercalation compounds.

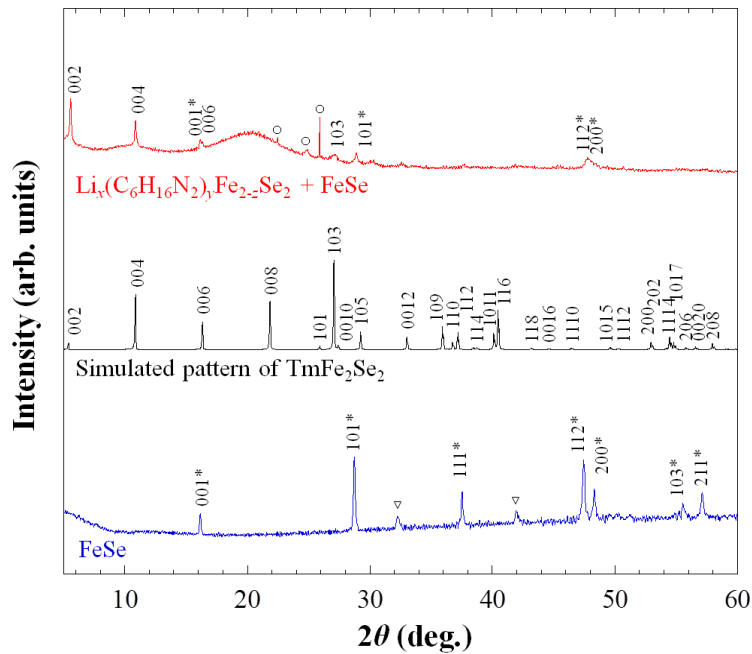

Fig. 1. (Color online) Powder x-ray diffraction pattern of the as-intercalated sample consisting of $Li_x(C_6H_{16}N_2)_yFe_{2-z}Se_2$ and FeSe, using $CuK_\alpha$ radiation. For reference, the simulated x-ray diffraction pattern of a hypothetical compound of $TmFe_2Se_2$ with the $ThCr_2Si_2$-type structure of $a = 3.453$ Å, $c = 32.450$ Å and the powder x-ray diffraction pattern of the host compound of FeSe are also shown. Indexes without and with asterisk are based on the $ThCr_2Si_2$-type and PbO-type structures, respectively. Peaks marked by ▽ and ○ are due to $Fe_7Se_8$ and $C_6H_{16}N_2$, respectively.

Table I. Lattice constants $a$ and $c$ of the host compound of FeSe, and FeSe and Li$_x$(C$_6$H$_{16}$N$_2$)$_y$Fe$_{2-z}$Se$_2$ in the as-intercalated sample. It is noted that the distance between the neighboring Fe layers, $d$, of Li$_x$(C$_6$H$_{16}$N$_2$)$_y$Fe$_{2-z}$Se$_2$ is given by the half of the $c$-axis leng.

|  | $a$ | $c$ |
|---|---|---|
| FeSe (host) | 3.765(1) | 5.510(3) |
| FeSe (as-intercalated sample) | 3.753(2) | 5.453(7) |
| Li$_x$(C$_6$H$_{16}$N$_2$)$_y$Fe$_{2-z}$Se$_2$ (as-intercalated sample) | 3.453(2) | 32.450(9) |

The intercalation of lithium and HMDA into FeSe has been confirmed by the chemical analysis using ICP-OES. The composition of the as-intercalated sample has been estimated as Li$_{1.25}$(C$_6$H$_{16}$N$_2$)$_{0.64}$Fe$_{2.05}$Se$_2$. Since some lithium and HMDA may exist in grain boundaries of the sample, and moreover, there remains a non-intercalated region of FeSe in the sample, however, the composition estimated by ICP-OES should be somewhat different from that of the region really including lithium and HMDA.

Figure 2 shows the TG curve on heating up to 900˚C at the rate of 1˚C/min for the as-intercalated sample consisting of $Li_x(C_6H_{16}N_2)_yFe_{2-z}Se_2$ and FeSe. Roughly, two steps of mass loss are observed; 21 % loss below 250˚C and large loss above 700˚C. The first step below 250˚C is inferred to be due to deintercalation or desorption of HMDA, because 21% of the mass of the as-intercalated sample is in good agreement with the weight percent of HMDA in the as-intercalated sample estimated by ICP-OES. The deintercalation of HMDA at low temperatures below 250˚C is very analogous to the deintercalation of EDA observed in $Li_x(C_2H_8N_2)_yFe_{2-z}Se_2$.[12]

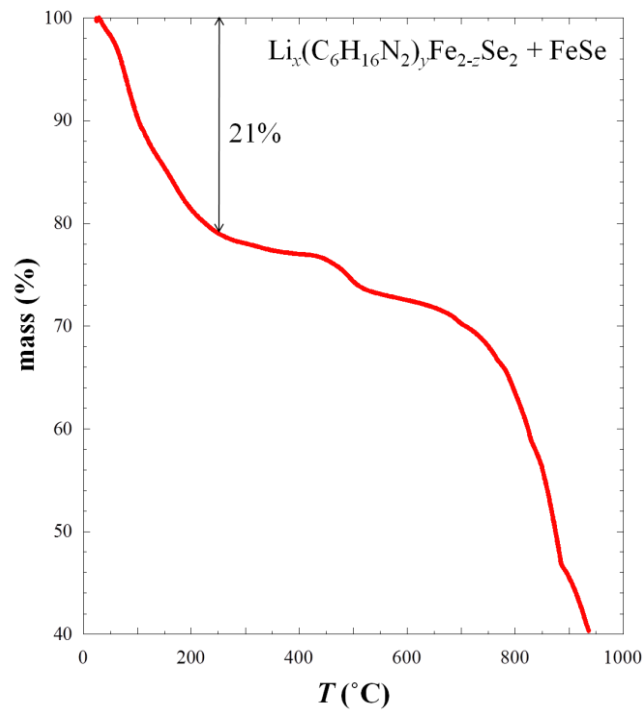

Fig. 2. Thermogravimetric (TG) curve on heating at the rate of 1˚C/min for the as-intercalated sample consisting of $Li_x(C_6H_{16}N_2)_yFe_{2-z}Se_2$ and FeSe.

Figure 3 shows the temperature dependence of $\chi$ in a magnetic field of 10 Oe on zero-field cooling (ZFC) and on field cooling (FC) for the as-intercalated sample consisting of $Li_x(C_6H_{16}N_2)_yFe_{2-z}Se_2$ and FeSe. The first superconducting transition is observed at 38 K and the second superconducting transition is faintly detected at 8 K. Taking into account the powder x-ray diffraction results, it is concluded that the first is due to bulk superconductivity of $Li_x(C_6H_{16}N_2)_yFe_{2-z}Se_2$, while the second is due to that of the non-intercalated region of FeSe. The superconducting volume fraction, simply estimated from the $\chi$ value at 2 K on ZFC, is ~ 9 %. It is noted that both the positive value of $\chi$ and the hysteresis of $\chi$ above $T_c$ seem to be due to magnetic impurities taken into the sample, as in the case of $Li_x(C_2H_8N_2)_yFe_{2-z}Se_2$ and $Li_xFe(Se,Te)$.[11,12,14]

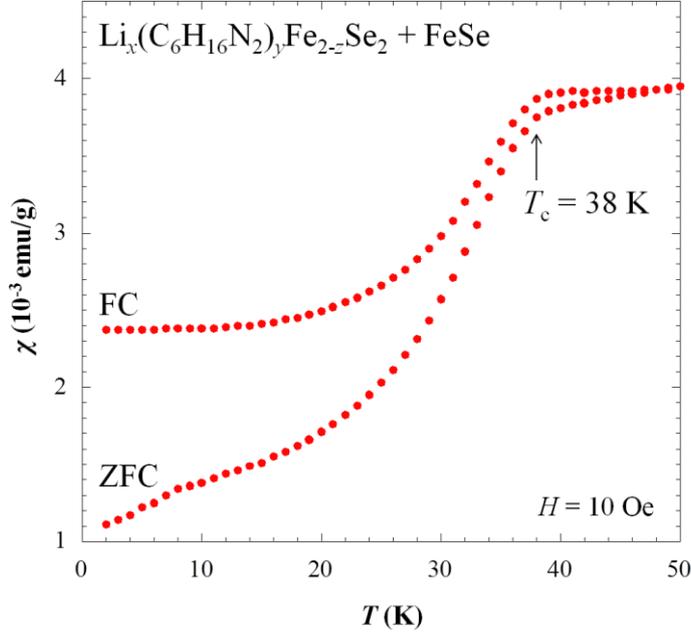

Fig. 3. Temperature dependence of the magnetic susceptibility, $\chi$, in a magnetic field of 10 Oe on zero-field cooling (ZFC) and field cooling (FC) for the as-intercalated sample consisting of $Li_x(C_6H_{16}N_2)_yFe_{2-z}Se_2$ and FeSe.

Figure 4 displays the temperature dependence of $\rho$ for the sintered (190˚C, 15 h) pellet sample consisting of $Li_x(C_6H_{16}N_2)_yFe_{2-z}Se_2$ and FeSe. It is found that $\rho$ shows a metallic temperature-dependence at high temperatures, though the value of $\rho$ is not so small probably due to insulating grain-boundaries in the sample. The superconducting transition is observed at low temperatures below ~ 44 K, though the transition is very broad and zero-resistivity is not observed above 5 K. The $T_c^{mid}$, defined as the temperature where $\rho$ shows a half of the normal-state value, is 18 K and much lower than $T_c$ = 38 K estimated from the $\chi$ measurements in the as-intercalated sample as described above. This is inferred to be due to the progress of the deintercalation of HMDA by the sintering at a temperature as high as 190˚C, taking into account the TG curve shown in Fig. 2. Similar effects of the sintering have been observed in $Li_x(C_2H_8N_2)_yFe_{2-z}Se_2$ also.[11] Here, it is noted that $T_c^{onset}$ is as high as ~ 44 K probably owing to the large superconducting fluctuation based on the two-dimensional electrical structure. Such a high $T_c^{onset}$ has been observed in sintered samples of $Li_x(C_2H_8N_2)_yFe_{2-z}Se_2$ also.[15]

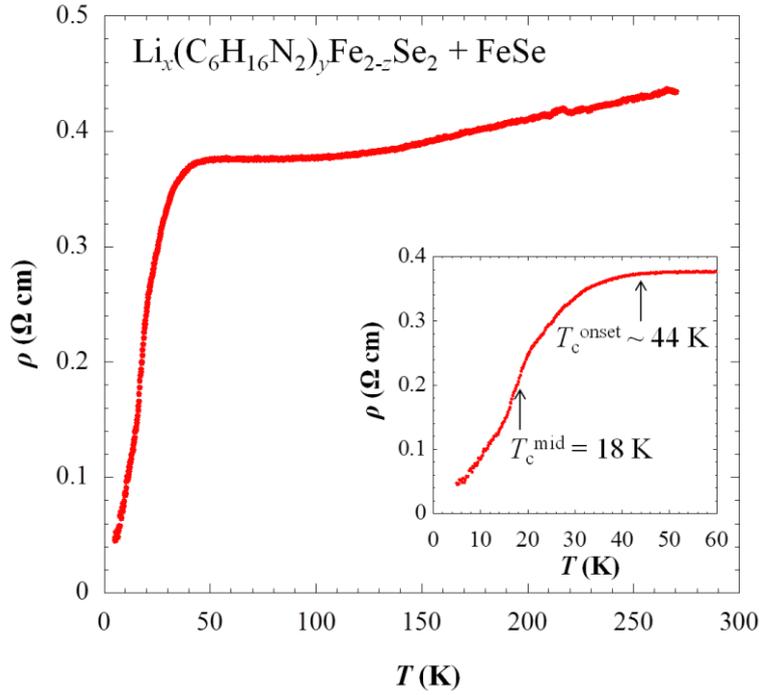

Fig. 4. Temperature dependence of the electrical resistivity, $\rho$, for the sintered (190˚C, 15 h) pellet sample consisting of $Li_x(C_6H_{16}N_2)_yFe_{2-z}Se_2$ and FeSe. The inset shows the temperature dependence of $\rho$ around $T_c$.

Finally, the maximum $T_c$'s of the FeSe-based intercalation superconductors obtained so far at ambient pressure are summarized in Fig. 5,[1,3-5,7-10,12,16-18] where their dependence on $d$ is plotted. It is found that $T_c$ increases monotonically with increasing $d$, is saturated at ~ 45 K and finally decreases. Such a domic dependence of $T_c$ on the interlayer spacing has been observed in alkali-earth-metal- and organic-molecule-intercalated HfNCl as well.[19] The increase in $T_c$ with increasing $d$ at $d \leq 9$ Å may be attributed to the enhancement of the pairing interaction on account of the possible improvement of the nesting condition at the Fermi surface owing to the enhancement of the two-dimensionality in the electrical structure.[20] However, the reason why $T_c$ is saturated and decreases with further increasing $d$ is not clear.

In summary, we have succeeded in synthesizing a new intercalation compound $Li_x(C_6H_{16}N_2)_yFe_{2-z}Se_2$ via intercalation of lithium and HMDA into FeSe. The $c$-axis length of $Li_x(C_6H_{16}N_2)_yFe_{2-z}Se_2$ is as long as 32.450(9) Å and therefore the $d$ value is 16.225(5) Å. These values are the largest among those of the FeSe-based intercalation compounds, indicating that not only lithium but also HMDA is intercalated into the sample. In fact, the composition of the as-intercalated sample has been estimated as

Li$_{1.25}$(C$_6$H$_{16}$N$_2$)$_{0.64}$Fe$_{2.05}$Se$_2$ by ICP-OES, though that of the region really including lithium and HMDA should be somewhat different. From the TG measurements, it has also been found that HMDA deintercalates from the sample at temperatures below 250˚C. Bulk superconductivity of Li$_x$(C$_6$H$_{16}$N$_2$)$_y$Fe$_{2-z}$Se$_2$ has been observed below 38 K in the $\chi$ measurements, and moreover, the resistive superconducting transition has been observed in the sintered pellet sample, though the transition is very broad due to the progress of the deintercalation of HMDA by the sintering. It has been found that the relation between $T_c$ and $d$ in the FeSe-based intercalation superconductors appears domic.

We would like to thank M. Ishikuro at Institute for Materials Research, Tohoku University, for his aid in the ICP-OES analysis.

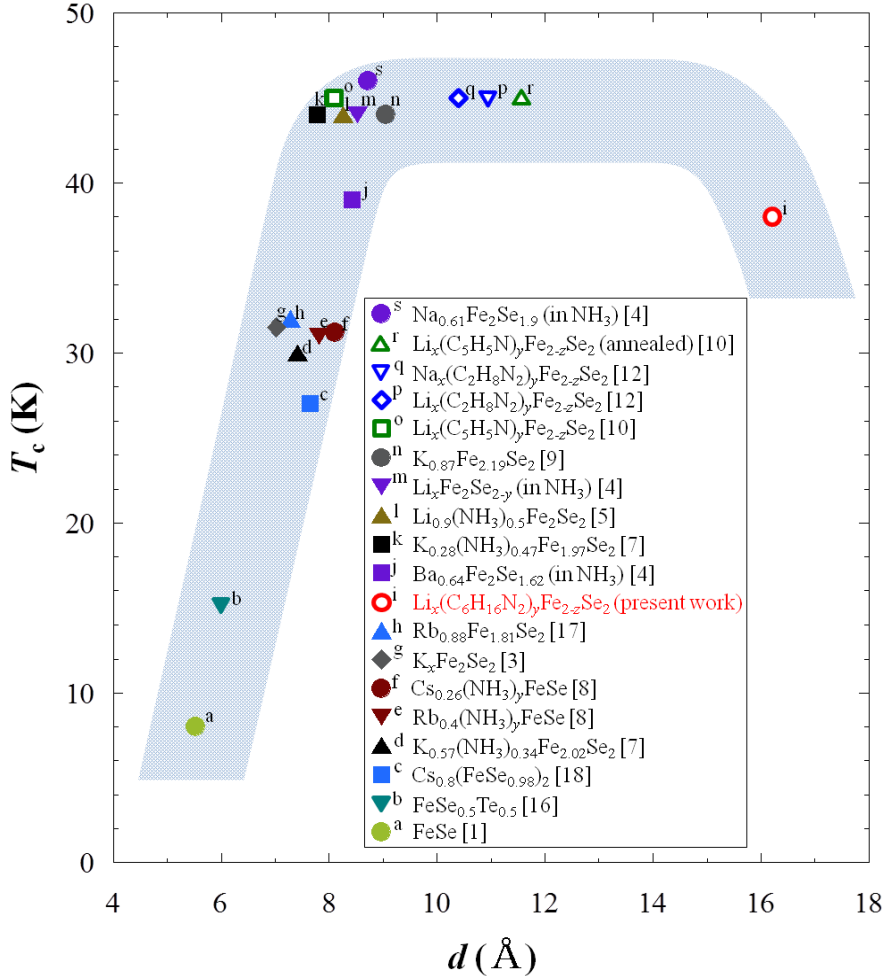

Fig. 5. (Color online) Relation between the maximum $T_c$ and the interlayer spacing, namely, the distance of the neighbouring Fe layers, $d$, in the FeSe-based intercalation superconductors.

**References**
1) F. -C. Hsu, J. -Y. Luo, K. -W. Yeh, T. -K. Chen, T. -W. Huang, P. M. Wu, Y. -C. Lee, Y. -L. Huang, Y. -Y. Chu, D. -C. Yan, and M. -K. Wu, Proc. Nati. Acad. Sci. **105**, 14262 (2008).
2) For a review, see G. R. Stewart, Rev. Mod. Phys. **83**, 1589 (2011).
3) J. Guo, S. Jin, G. Wang, S. Wang, K. Zhu, T. Zhou, and M. He, X. Chen, Phys. Rev. B **82**, 180520(R) (2010).
4) T. P. Ying, X. L. Chen, G. Wang, S. F. Jin, T. T. Zhou, X. F. Lai, H. Zhang, and W. Y. Wang, Sci. Rep. **2**, 426 (2012).
5) E. -W. Scheidt, V. R. Hathwar, D. Schmitz, A. Dunbar, W. Scherer, F. Mayr, V. Tsurkan, J. Deisenhofer, and A. Loidl, Eur. Phys. J. B **85**, 279 (2012).
6) M. Burrard-Lucas, D. G. Free, S. J. Sedlmaier, J. D. Wright, S. J. Cassidy, Y. Hara, A. J. Corkett, T. Lancaster, P. J. Baker, S. J. Blundell, and S. J. Clarke, Nature Mater. **12**, 15 (2013).
7) T. Ying, X. Chen, G. Wang, S. Jin, X. Lai, T. Zhou, H. Zhang, S. Shen, and W.Wang, J. Am. Chem. Soc. **135**, 2951 (2013).
8) L. Zheng, M. Izumi, Y. Sakai, R. Eguchi, H. Goto, Y. Takabayashi, T. Kambe, T. Onji, S. Araki, T. C. Kobayashi, J. Kim, A. Fujiwara, and Y.Kubozono, Phys. Rev. B **88**, 094521 (2013).
9) A. -M. Zhang, T. -L. Xia, K. Liu, W. Tong, Z. -R. Yang, and Q. -M. Zhang, Sci. Rep. **3**, 1216 (2013).
10) A. Krzton-Maziopa, E. V. Pomjakushina, V. Y. Pomjakushin, F. Rohr, A. Schilling, and K. Conder, J. Phys.: Condens. Matter **24**, 382202 (2012).
11) T. Hatakeda, T. Noji, T. Kawamata, M. Kato, and Y. Koike, J. Phys. Soc. Jpn. **82**, 123705 (2013).
12) T. Noji, T. Hatakeda, S. Hosono, T. Kawamata, M. Kato, and Y. Koike, Physica C **504**, 8 (2014).
13) F. Izumi and K. Momma, Solid State Phenom. **130**, 15 (2007).
14) H. Abe, T. Noji, M. Kato, and Y. Koike, Physica C **470**, S487 (2010).
15) T. Hatakeda, T. Noji, S. Hosono, T. Kawamata, M. Kato, and Y. Koike, to be published.
16) K. -W. Yeh, T. -W. Huang, Y. -L. Huang, T. -K. Chen, F. -C. Hsu, P. M. Wu, Y. -C. Lee, Y. -Y. Chu, C. -L. Chen, J. -Y. Luo, D. -C. Yan, and M. -K. Wu, Europhys. Lett. **84**, 37002 (2008).

17) A. F. Wang, J. J. Ying, Y. J. Yan, R. H. Liu, X. G. Luo, Z. Y. Li, X. F. Wang, M. Zhang, G. J. Ye, P. Cheng, Z. J. Xiang, and X. H. Chen, Phys. Rev. B **83**, 060512(R) (2011).
18) A. Krzton-Maziopa, Z. Shermadini, E. Pomjakushina, V. Pomjakushin, M. Bendele, A. Amato, R. Khasanov, H. Luetkens, and K. Conder, J. Phys.: Condens. Matter **23**, 052203 (2011).
19) S. Zhang, M. Tanaka, H. Zhu, and S. Yamanaka, Supercond. Sci. Technol. **26**, 085015 (2013).
20) T. Takano, T. Kishiume, Y. Taguchi, and Y. Iwasa, Phys. Rev. Lett. **100**, 247005 (2008).